\documentclass[amsmath,amssymb,superscriptaddress,nobalancelastpage,prb,twocolumn]{revtex4-1}

\usepackage{bm}
\usepackage{braket}
\usepackage{diagbox}
\usepackage{graphicx}
\usepackage{hyperref}
\hypersetup{colorlinks,linkcolor=blue,urlcolor=blue,citecolor=blue}
\usepackage{siunitx}
\usepackage{ulem}
\usepackage{varioref}
\usepackage{xr-hyper}
\usepackage{xcolor}

\newcommand{\cyan}[1]{{\textcolor{cyan}{#1}}}

\newcommand{\LSCO}{La$_{2-x}$Sr$_{x}$CuO$_4$}
\newcommand{\PLCCO}{Pr$_{1.3-x}$La$_{0.7}$Ce$_{x}$CuO$_4$}
\newcommand{\NCCO}{Nd$_{2-x}$Ce$_{x}$CuO$_4$}
\newcommand{\CSRO}{Ca$_{1.8}$Sr$_{0.2}$RuO$_4$}

\newcommand{\dxz}{$d_{xz}$}
\newcommand{\dyz}{$d_{yz}$}

\newcommand{\dx}{$d_{x^2-y^2}$}

\begin{document}

 \author{M.~Horio}
 \email{mhorio@issp.u-tokyo.ac.jp}
   \affiliation{Physik-Institut, Universit\"{a}t Z\"{u}rich, Winterthurerstrasse 190, CH-8057 Z\"{u}rich, Switzerland}
   
     \author{K.~P.~Kramer}
     \affiliation{Physik-Institut, Universit\"{a}t Z\"{u}rich, Winterthurerstrasse 190, CH-8057 Z\"{u}rich, Switzerland}
   
   \author{Q.~Wang}
     \affiliation{Physik-Institut, Universit\"{a}t Z\"{u}rich, Winterthurerstrasse 190, CH-8057 Z\"{u}rich, Switzerland}
   
    \author{A.~Zaidan}
     \affiliation{Physik-Institut, Universit\"{a}t Z\"{u}rich, Winterthurerstrasse 190, CH-8057 Z\"{u}rich, Switzerland}
   
   \author{K.~von~Arx}
     \affiliation{Physik-Institut, Universit\"{a}t Z\"{u}rich, Winterthurerstrasse 190, CH-8057 Z\"{u}rich, Switzerland}
   
    \author{D.~Sutter}
     \affiliation{Physik-Institut, Universit\"{a}t Z\"{u}rich, Winterthurerstrasse 190, CH-8057 Z\"{u}rich, Switzerland}
   
     \author{C.~E.~Matt}
     \affiliation{Physik-Institut, Universit\"{a}t Z\"{u}rich, Winterthurerstrasse 190, CH-8057 Z\"{u}rich, Switzerland}
   
     \author{Y.~Sassa}
  \affiliation{Department of Physics, Chalmers University of Technology, SE-412 96 G\"{o}teborg, Sweden}

\author{N.~C.~Plumb}
 \affiliation{Swiss Light Source, Paul Scherrer Institut, CH-5232 Villigen PSI, Switzerland}
 
\author{M.~Shi}
 \affiliation{Swiss Light Source, Paul Scherrer Institut, CH-5232 Villigen PSI, Switzerland}
 
 \author{A.~Hanff}
 \affiliation{Institut f\"ur Experimentelle und Angewandte Physik, Christian-Albrechts-Universit\"at zu Kiel, D-24098 Kiel, Germany}
 \affiliation{Ruprecht Haensel Laboratory, Christian-Albrechts-Universit\"at zu Kiel, D-24098 Kiel, Germany}
 
  \author{S.~K.~Mahatha}
 \affiliation{Ruprecht Haensel Laboratory, Deutsches Elektronen-Synchrotron DESY, D-22607 Hamburg, Germany}
 
 \author{H.~Bentmann}
 \affiliation{Experimentelle Physik VII and W\"urzburg-Dresden Cluster of Excellence ct.qmat, Universit\"at W\"urzburg, Am Hubland, D-97074 W\"urzburg, Germany}
 
  \author{F.~Reinert}
 \affiliation{Experimentelle Physik VII and W\"urzburg-Dresden Cluster of Excellence ct.qmat, Universit\"at W\"urzburg, Am Hubland, D-97074 W\"urzburg, Germany}
 
  \author{S.~Rohlf}
 \affiliation{Institut f\"ur Experimentelle und Angewandte Physik, Christian-Albrechts-Universit\"at zu Kiel, D-24098 Kiel, Germany}
 \affiliation{Ruprecht Haensel Laboratory, Christian-Albrechts-Universit\"at zu Kiel, D-24098 Kiel, Germany}
 
 \author{F.~K.~Diekmann}
 \affiliation{Institut f\"ur Experimentelle und Angewandte Physik, Christian-Albrechts-Universit\"at zu Kiel, D-24098 Kiel, Germany}
 \affiliation{Ruprecht Haensel Laboratory, Christian-Albrechts-Universit\"at zu Kiel, D-24098 Kiel, Germany}
 
  \author{J.~Buck}
 \affiliation{Institut f\"ur Experimentelle und Angewandte Physik, Christian-Albrechts-Universit\"at zu Kiel, D-24098 Kiel, Germany}
 \affiliation{Ruprecht Haensel Laboratory, Deutsches Elektronen-Synchrotron DESY, D-22607 Hamburg, Germany}
  
    \author{M.~Kall\"ane}
 \affiliation{Institut f\"ur Experimentelle und Angewandte Physik, Christian-Albrechts-Universit\"at zu Kiel, D-24098 Kiel, Germany}
  \affiliation{Ruprecht Haensel Laboratory, Christian-Albrechts-Universit\"at zu Kiel, D-24098 Kiel, Germany}
  
  \author{K.~Rossnagel}
 \affiliation{Institut f\"ur Experimentelle und Angewandte Physik, Christian-Albrechts-Universit\"at zu Kiel, D-24098 Kiel, Germany}
  \affiliation{Ruprecht Haensel Laboratory, Deutsches Elektronen-Synchrotron DESY, D-22607 Hamburg, Germany}

 \author{E.~Rienks}
 \affiliation{Helmholtz Zentrum Berlin, Bessy II, D-12489 Berlin, Germany}

   \author{V.~Granata}
\affiliation{CNR-SPIN, I-84084 Fisciano, Salerno, Italy}
\affiliation{Dipartimento di Fisica "E.R.~Caianiello", Universit\`{a} di Salerno, I-84084 Fisciano, Salerno, Italy}
 
 \author{R.~Fittipaldi}
\affiliation{CNR-SPIN, I-84084 Fisciano, Salerno, Italy}
\affiliation{Dipartimento di Fisica "E.R.~Caianiello", Universit\`{a} di Salerno, I-84084 Fisciano, Salerno, Italy}

 \author{A.~Vecchione}
\affiliation{CNR-SPIN, I-84084 Fisciano, Salerno, Italy}
\affiliation{Dipartimento di Fisica "E.R.~Caianiello", Universit\`{a} di Salerno, I-84084 Fisciano, Salerno, Italy}
 
  \author{T.~Ohgi}
  \affiliation{Department of Applied Physics, Tohoku University, Sendai 980-8579, Japan}
  
  \author{T.~Kawamata}
  \affiliation{Department of Applied Physics, Tohoku University, Sendai 980-8579, Japan}
 
  \author{T.~Adachi}
  \affiliation{Department of Engineering and Applied Sciences, Sophia University, Tokyo 102-8554, Japan}
  
  \author{Y.~Koike}
  \affiliation{Department of Applied Physics, Tohoku University, Sendai 980-8579, Japan}
  
  \author{A.~Fujimori}
  \affiliation{Department of Physics, University of Tokyo, Bunkyo-ku, Tokyo 113-0033, Japan}
  \affiliation{Department of Applied Physics, Waseda University, Shinjuku, Tokyo 169-8555, Japan}
   
    \author{M.~Hoesch}
 \affiliation{Deutsches Elektronen-Synchrotron DESY, Photon Science, Notkestrassse 85, 22607 Hamburg, Germany}
   
\author{J.~Chang}
\email{johan.chang@physik.uzh.ch}
    \affiliation{Physik-Institut, Universit\"{a}t Z\"{u}rich, Winterthurerstrasse 190, CH-8057 Z\"{u}rich, Switzerland}


 
   \title{Fermi Liquid Universality Revealed by Electron Spectroscopy} 
   
    \title{\cyan{Isotropic Fermi Liquid in Electron-Overdoped \PLCCO}} 
    
    \title{Oxide Fermi liquid universality revealed by electron spectroscopy}
   
\begin{abstract}
We present a combined soft x-ray and high-resolution vacuum-ultraviolet angle-resolved photoemission spectroscopy study of the electron-overdoped cuprate \PLCCO\ (PLCCO). Demonstration of its highly two-dimensional band structure enabled precise determination of the in-plane self-energy dominated by electron-electron scattering. Through analysis of this self-energy and the
Fermi-liquid cut-off energy scale, 
we find -- in contrast to hole-doped cuprates -- a momentum isotropic and comparatively weak electron correlation in PLCCO.
Yet, the self-energies extracted from multiple oxide systems combine 
to demonstrate
a logarithmic divergent relation between the quasiparticle scattering rate and mass.
This 
constitutes a spectroscopic version of the Kadowaki-Woods relation with an important  merit -- the demonstration of Fermi liquid quasiparticle lifetime and mass being set by a single energy scale.

\end{abstract}

\maketitle

\section{Introduction}   
The Fermi liquid quasiparticle concept underpins much of our understanding of correlated metals~\cite{LandauFL,Coleman2001,NevenPNAS2013,DoironLeyraud07a}. Electron-electron interaction renormalizes the quasiparticle lifetime and mass whereas spin and charge quantum numbers are identical to the non-interacting limit. 
This quasiparticle identity 
assures electronic specific heat $C$ to scale with temperature $T$ and a resistivity proportional to $T^2$ below an energy scale $\omega_c$. In the limit $k_B T\ll \omega_c$, the 
Wiedemann-Franz law~\cite{WiedemannFranz} dictates a fundamental relation between heat and charge conduction. Under sufficiently strong electron correlation ($\omega_c\rightarrow 0$), the Fermi liquid 
breaks down and is replaced by a Mott insulating or non-Fermi liquid state. Studying this breakdown route is an important step to conceptualize non-Fermi liquids that are often found in the context of unconventional superconductivity~\cite{CooperScience2009,KasaharaPRB2010,JinNat2011,PfauNature2012}.

The Kadowaki-Woods relation~\cite{RicePRL1968,KadowakiSSC1986} suggests a link between Fermi-liquid quasiparticle lifetime and mass renormalization. The resistivity coefficient $A$ in $\rho=A T^2$ 
reflects a momentum integrated lifetime whereas the Sommerfeld coefficient $\gamma$ -- inferred from specific heat -- yields the mass. Accumulated 
empirical evidence supports the Kadowaki-Woods proposal of the ratio $A/ \gamma^2$ being invariant with respect to the electron-electron interaction strength~\cite{RicePRL1968,KadowakiSSC1986}. The Kadowaki-Woods ratio also has a theoretical foundation starting from the electronic self-energy~\cite{MiyakeSSP1989,JackoNatPhys2009}. 
Although photoemission spectroscopy has angle (momentum) resolving capability and direct access to the self-energy, 
no spectroscopic 
evidence of the Kadowaki-Woods relation has been established. This lack of progress stems from a chain of challenges: (i) Photoemission spectroscopy is best suited for two-dimensional systems~\cite{DamascelliRMP03}, narrowing down the range of studiable materials. (ii) Self-energy analysis of quasi two-dimensional systems 
is limited by residual $k_z$ and disorder
broadening~\cite{StrocovJESRP2003} in the weak coupling limit. (iii) The strong coupling limit leads to energy scales below the resolving power.

\begin{figure*}[ht]
 	\begin{center}
 		\includegraphics[width=0.99\textwidth]{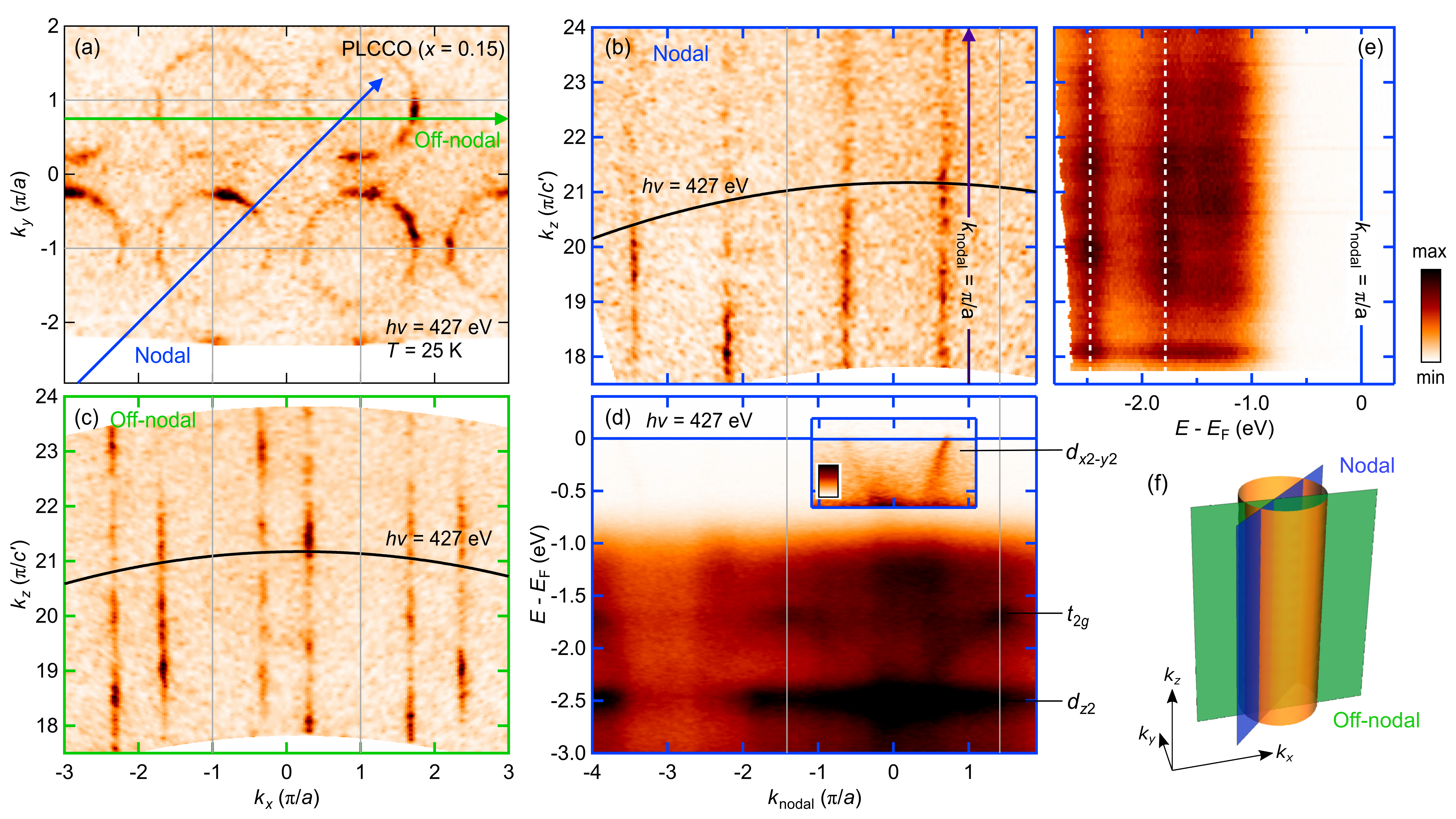}
 	\end{center}
 	\caption{\textbf{Two-dimensional Fermi surface of PLCCO ($\bm{x = 0.15}$).} (a) In-plane Fermi surface, for the integration window $E_\mathrm{F} \pm \SI{30}{meV}$,
 	measured at $T=\SI{25}{K}$ and $h\nu = \SI{427}{eV}$. 
 	(b),(c) Out-of-plane Fermi surface maps recorded along the 
 	nodal and off-nodal directions, respectively, as indicated in (a). (d) Energy distribution map taken along the nodal direction. 
 	Cu $d$-orbital characters are assigned 
 	as in Ref.~\onlinecite{KramerPRB2019}. Gray lines in (a)--(d) indicate Brillouin-zone boundaries. (e) Energy distribution map taken along $k_z$ for
 	a fixed in-plane momentum -- 
 	$k_\mathrm{nodal} = \pi/a$. White dashed lines are guides to the eye. (f) Schematic Fermi surface with the nodal and off-nodal cuts.}
 	\label{fig:fig1}
 \end{figure*}

Here, we demonstrate by soft x-ray (SX) angle-resolved photoemission spectroscopy (ARPES) that the electron-overdoped cuprate superconductor \PLCCO\ (PLCCO) has a two-dimensional electronic structure with negligible $k_z$ dispersion. This result justifies and enables extraction of the in-plane self-energy using vacuum-ultraviolet (VUV) ARPES. In contrast to hole-overdoped \LSCO\ (LSCO), which hosts non-local interactions~\cite{ChangNatComm13,RayJPCM2019}, an essentially momentum-isotropic Fermi-liquid self-energy is found from the nodal to antinodal region. Again in direct comparison to LSCO, much weaker electron-electron interactions are observed in PLCCO. This result is reflected both in 
quasiparticle lifetime and the Fermi liquid cut-off energy scale linked to the mass renormalization factor $Z$. Combined with results on other correlated (non-superconducting) oxide systems, these results sum into a spectroscopic version of the Kadowaki-Woods relation where the quasiparticle scattering rate $\beta$ scales with $Z^{-2}$ over more than an order of magnitude.
This relation connects weakly and strongly correlated Fermi liquids via a single energy scale.

\section{Methods} 
Single crystals of PLCCO with $x=0.15$ were synthesized by the traveling-solvent floating-zone method. 
After reduction annealing~\cite{AdachiJPSJ2013,HorioNatCommun2016} at \SI{800}{\celsius} for \SI{24}{h}, the 
overdoped sample showed superconductivity with $T_c=\SI{19}{K}$ -- lower than the optimal $T_c\sim \SI{27}{K}$~\cite{HorioNatCommun2016}. 
The quality of our crystal is reflected by a residual resistivity $\rho_0= 38$ $\mu\Omega$cm. 
SX and VUV ARPES experiments were carried out at the P04 and Surface/Interface Spectroscopy (SIS) beamline at DESY and Swiss Light Source~\cite{SIS}, respectively. Samples were
cleaved \textit{in situ} under ultra high vacuum ($< \SI{5e-11}{Torr}$) by employing a
top-post method. Circularly polarized incident photons of $h\nu = 30$ -- \SI{600}{eV} were used for both experiments. The effective energy resolution (temperature) was set to $\sim \SI{50}{meV}$ (\SI{25}{K}) for the SX and \SIrange{14}{17}{meV} (\SI{18}{K}) for the VUV measurements. For both setups, the angular resolution is $\sim$0.15 degrees. 

 \begin{figure*}[ht]
 	\begin{center}
 		\includegraphics[width=0.99\textwidth]{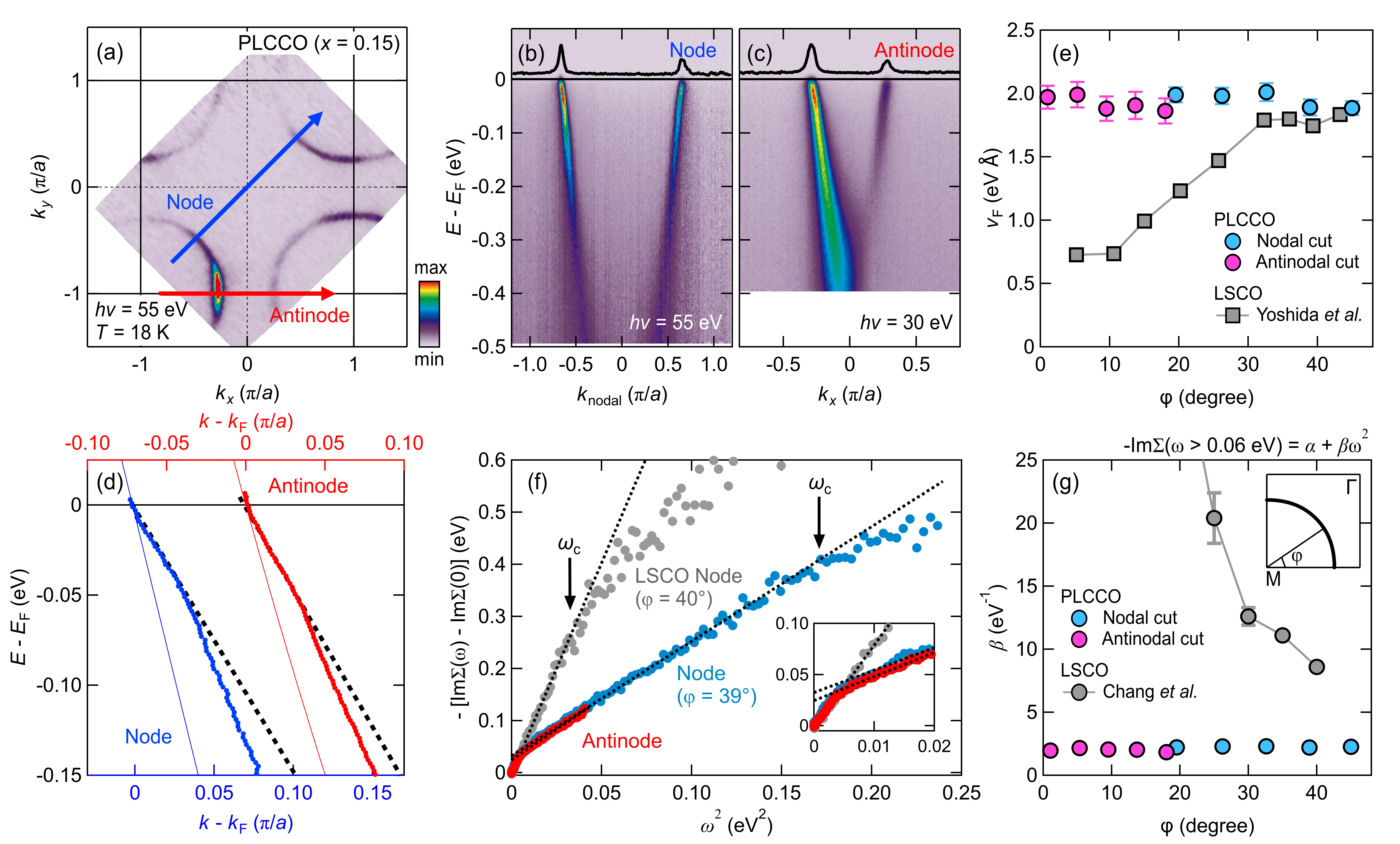}
 	\end{center}
 	\caption{\textbf{In-plane isotropic 
 	self-energy structure of PLCCO.} (a) Fermi surface recorded at the indicated temperature and photon energy. 
 	(b),(c) Nodal and antinodal energy distribution maps. 
 Different photon energies are used to enhance the matrix element. 
 Solid black lines are the momentum distribution curves (MDCs) at the Fermi level. The MDC linewidth and Fermi velocities yield Im$\Sigma(0)\sim$ 0.077~eV and 0.085~eV for respectively the nodal and antinodal  direction. 
 	(d) Nodal and antinodal band dispersions extracted from MDC analysis. Solid curves and dashed lines represent bare bands and extrapolation of low-energy dispersions, respectively.
 	(e) Fermi velocity $v_\mathrm{F}$ of PLCCO and LSCO ($x=0.22$)~\cite{YoshidaJCMP07} plotted as a function of the Fermi surface angle $\varphi$ [see inset of (g)].
 	 (f) Near nodal self-energy 
 	 -Im$\Sigma$($\omega$) plotted versus 
 	 $\omega^2$ for PLCCO and LSCO ($x=0.23$)~\cite{ChangNatComm13}.
 	 Dotted curves are fits revealing the $- \mathrm{Im} \Sigma(\omega) \propto \beta \omega^2$ dependence and 
 	 black arrows mark high-energy deviation (Fermi liquid cut-off). Low-energy part is magnified in the inset.
 	 (g) Coefficient $\beta$ for the $\omega^2$ term of Im$\Sigma$($\omega$) for PLCCO and LSCO ($x=0.23$)~\cite{ChangNatComm13}.
 	}	
	 	\label{fig:fig2}
 \end{figure*}

\section{Results}
Using SX-ARPES, which provides  
comparatively good $k_z$ resolution~\cite{StrocovJESRP2003}, we evaluate the dimensionality of the electronic structure in PLCCO. Along the nodal and off-nodal cuts [see in-plane Fermi surface map in Fig.~\ref{fig:fig1}(a)], the Fermi surface was investigated in the $k_z$ direction over three Brillouin zones [Figs.~\ref{fig:fig1}(b) and (c)]. 
Within 
the experimental resolution, the Fermi surface (with \dx\ character) has no $k_z$ dispersion. 
Consistently, none of the 
$d$ bands ($t_{2g}$ and $d_{z^2}$) 
at deeper binding energies [Fig.~\ref{fig:fig1}(d)]~\cite{KramerPRB2019} exhibits any significant dispersions along the $k_z$ direction [Fig.~\ref{fig:fig1}(e)]. These highly two-dimensional characteristics of PLCCO are in 
contrast to the recently unveiled three-dimensional electronic structure of the hole-overdoped cuprate LSCO~\cite{MattNatCommun2018,HorioPRL2018}. 
This difference stems from a 
reduced inter-layer hopping due to the absence of apical oxygen atoms in the electron-doped cuprates~\cite{PhotopoulosANDP2019}. The two-dimensional nature of the electron-doped cuprates is also reflected by a large resistivity anisotropy $\rho_c / \rho_{ab} > 10000$~\cite{OnosePRB2004}.
This is ten and hundred times larger than the anisotropies reported in Sr$_2$RuO$_4$~\cite{HusseyPRB1998} and overdoped LSCO~\cite{NakamuraPRB1993}, respectively. 


The established two-dimensional electronic structure
 of PLCCO justifies use of surface-sensitive VUV light for extraction of the self-energy.
The Fermi surface recorded 
at $h\nu = 55$~eV [Fig.~\ref{fig:fig2}(a)] -- essentially identical to that observed with SX [Fig.~\ref{fig:fig1}(a)] -- corresponds to a filling of 15~\% electron doping. While there have been extensive reports on additional electron doping by reduction annealing of electron-doped cuprates ~\cite{HorioNatCommun2016,WeiPRL2016,Song2017,HorioPRL2018NCO,HorioPRB2018,AsanoJPSJ2018}, this filling is
consistent with the nominal Ce concentration.

The two-dimensional Fermi surface and the absence of (i) hot spots~\cite{ArmitagePRL2001} and (ii) van Hove singularities near the Fermi level 
form the basis for 
self-energy analysis across the
entire Brillouin zone.
Low-energy quasiparticle excitations were recorded along 
nodal and antinodal directions [see Fig.~\ref{fig:fig2}(a)].
Nodal and antinodal energy distribution maps shown in Figs.~\ref{fig:fig2}(b) and (c), taken respectively with $h\nu = 55$ and \SI{30}{eV} incident light, reveal sharp and dispersive quasiparticle peaks. 
In agreement with previous studies~\cite{SchmittPRB2008,ParkPRL2008,Liu2012}, both dispersions exhibit (possibly electron-phonon coupled) 
kinks at the binding energy of $\sim \SI{0.05}{eV}$ [Fig.~\ref{fig:fig2}(d)].
Fermi velocities $v_\mathrm{F}$ -- plotted as a function of the Fermi surface angle $\varphi$ in Fig.~\ref{fig:fig2}(e) -- 
are extracted by fitting the quasiparticle dispersion up to the kink energy scale. 
In contrast to the strongly anisotropic $v_\mathrm{F}$ in overdoped LSCO ($x=0.22$)~\cite{YoshidaJCMP07}, $v_\mathrm{F}$ is found to be almost independent of momentum in PLCCO. This marked difference is 
linked to the 
proximity of the van Hove singularity to the Fermi level in LSCO~\cite{YoshidaPRB2006}. 

 \begin{figure}[ht!]
 	\begin{center}
 		\includegraphics[width=0.45\textwidth]{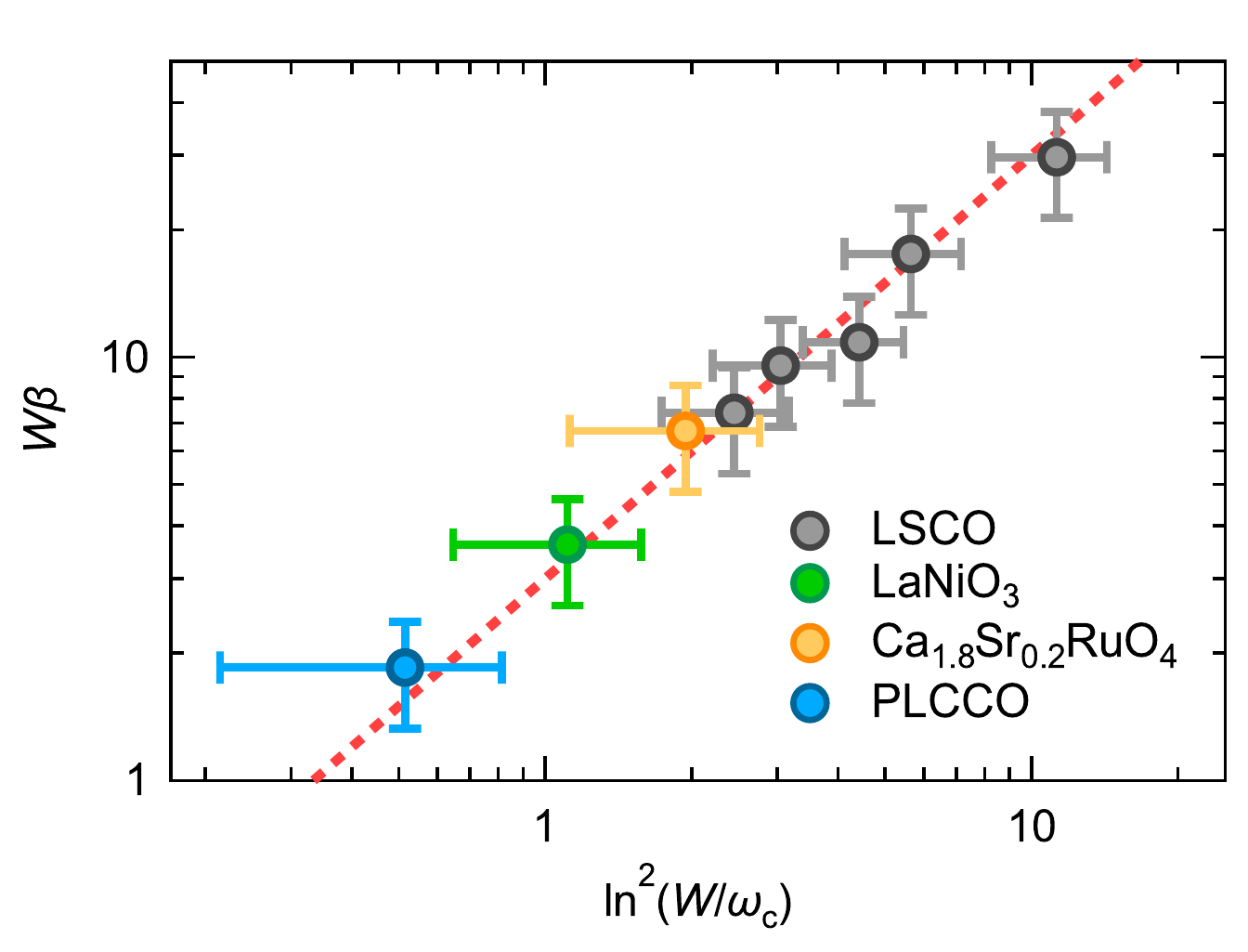}
 	\end{center}
 	\caption{\textbf{Spectroscopic Kadowaki-Woods relation.} 
 	$W\beta$ versus  ln$^2 (W / \omega_c)$, where $\beta$ is the prefactor in $- \mathrm{Im} \Sigma(\omega) \propto\beta \omega^2$, $\omega_c$ is the Fermi-liquid cut-off energy, and $W$ is a bare energy scale for LSCO ($x=0.23$)~\cite{ChangNatComm13}, LaNiO$_3$~\cite{EguchiPRB2009}, \CSRO, and PLCCO ($x=0.15$). The LSCO data points stem from different Fermi momenta. Red dashed line is a linear fit of the plotted data. 
 	Error bars are set by assuming 20~\% of uncertainty on the used bare-band velocities.
 	}	
	 	\label{fig:fig3}
 \end{figure}

ARPES spectra contain information about the electronic self-energy $\Sigma(k,\omega)$ through its relation to the spectral function $A(k,\omega) = -1/\pi$ $\mathrm{Im}[1/(\omega- \varepsilon_k - \Sigma(k,\omega))]$ where $\varepsilon_k$ is the bare band dispersion. 
The quasiparticle lifetime 
is 
obtained through Im$\Sigma(k,\omega) = v_k \Gamma_k$ where $v_k = \partial \varepsilon_k / \partial k$ is the bare band velocity and $\Gamma_k$ is momentum-distribution-curve (MDC) half width at half maximum~\cite{ChangPRB08,VallaScience1999,MeevasanaPRB2008}.
In our case, the nodal MDC linewidth yields a mobility $\mu=e/(\hbar k_F \Gamma_k)=14.3$~cm$^2$/Vs (see Ref.~\onlinecite{YoshidaPRB2009}) consistent with that inferred from transport $\mu=(ne\rho_0)^{-1}=13.9$~cm$^2$/Vs using $n=1.15$ per Cu atom and residual resistivity $\rho_0= 38$ $\mu\Omega$cm.
To estimate the bare band velocity, we 
fit the Fermi surface to the following single-band tight-binding model: 
\begin{align}
\epsilon &= \epsilon _0 - 2t (\mathrm{cos} \ k_xa + \mathrm{cos} \ k_ya) \nonumber \\
&-4t'\mathrm{cos} \ k_xa \ \mathrm{cos} \ k_ya -2t''(\mathrm{cos} \ 2k_xa + \mathrm{cos} \ 2k_ya),
\end{align}
which includes nearest ($t$), second-nearest ($t'$), and third-nearest ($t''$) neighbor hopping parameters. With  $\epsilon_0$ being the 
 band center, we find $\epsilon_0/t = -0.04$ and $t'/t = -0.19$ when using $t''/t'=-1/2$. 
Assuming $t = \SI{0.41}{eV}$ based on a previous density-functional-theory (DFT) estimate on \NCCO~\cite{IkedaPRB2009}, the full two-dimensional bare
band structure
is constructed. 
This enables extraction of the self-energy 
Im$\Sigma(k,\omega)$ 
as 
illustrated for 
cuts through node and antinode in Fig.~\ref{fig:fig2}(f).
ARPES spectra were recorded up to a binding energy of \SI{0.5}{eV}. However, the antinodal Im$\Sigma(\omega)$ is plotted only for 
$\omega^2 < \SI{0.04}{eV^2}$
as analysis above this energy scale is challenged by the van Hove singularity~\cite{SchmittPRB2011}.

\section{Discussion}
Both the nodal and antinodal Im$\Sigma(\omega)$ curves display a kink at $\omega \sim \SI{0.06}{eV}$ [see inset of Fig.~\ref{fig:fig2}(f)], 
Kramers-Kronig consistently with the kink observed in the band dispersion. Below this phonon cut-off energy scale $\omega_{ph}\approx 0.06$~eV, the self-energy is expected to contain contributions from both electron-phonon and electron-electron interactions. Probing in the $\omega\rightarrow 0$ limit allows -- in principle -- direct comparison to low-temperature transport properties~\cite{JinNat2011,IkedaPRB2016,SarkarPRB2018}.
With our experimental temperature and energy resolution, however, 
we cannot distinguish 
a Fermi liquid with Im$\Sigma \propto \omega^2$ from, for example, a marginal Fermi liquid with Im$\Sigma \propto \sqrt{\omega^2 + (\pi k_\mathrm{B}T)^2}$~\cite{AbrahamsPNAS}. Excitations observed above the 
kink 
energy scale (\SI{0.06}{eV}) do not 
pose these limitations and hence offers direct insight into the electron-electron interactions. .  
As electron-phonon self-energy contribution saturates for $\omega>\omega_{ph}$, 
the electron-phonon interactions is effectively filtered out of the analysis.
Furthermore, our energy resolution does not limit the analysis of the quasiparticle excitations in this regime. 
The extracted electron-electron interacting self-energy 
is parametrized by
$-$Im$\Sigma(\omega)=\alpha + \beta \omega^2$, with $\alpha$ and $\beta$ being constants. 
This parameterization implicitly assumes that different scattering channels (electron-disorder, electron-phonon and electron-electron etc.) are additive. A similar premise is used for analysis of resistivity measurements on related electron-doped cuprates~\cite{JinNat2011}.
As demonstrated in Fig.~\ref{fig:fig2}(f), this parabolic function convincingly fits the Im$\Sigma(\omega)$ curves 
over a wide energy range ($0.06<\omega < \SI{0.4}{eV}$). 

This 
functional form of the self-energy is identical to a 
three-dimensional Fermi-liquid 
which displays Im$\Sigma(\omega) -$Im$\Sigma(0) = -\beta \omega^2$ below a cut-off energy $\omega_c$~\cite{MiyakeSSP1989,ByczukNatPhys2007,JackoNatPhys2009,ChangNatComm13}. In two dimensions, a logarithmic correction~\cite{InglePRB2005} influences mostly the self-energy for $\omega\ll \varepsilon_\mathrm{F} \sim 1.5$~eV~\cite{KramerPRB2019} and an approximate Im$\Sigma(\omega)\propto\beta'\omega^2$ dependence remains in the considered $\omega$ 
range while $\beta'$ is weakly overestimating $\beta$. The coefficient $\beta=\lambda/\omega_c^2$ -- given by the bare scattering rate $\lambda$ and $\omega_c$ -- reflects the effective electron-electron interaction strength.
In Fig.~\ref{fig:fig2}(g), $\beta$ -- plotted versus Fermi surface angle -- appears essentially isotropic (momentum independent). 
This is in strong contrast to the hole-overdoped counterpart LSCO where $\beta$ is 
highly anisotropic and takes on much larger values already in the nodal region [see Fig.~\ref{fig:fig2}(g)]. This weaker electron correlation strength found for electron-overdoped cuprates is consistent with theoretical proposals~\cite{SenechalPRL2004,DasPRB2009,WeberNatPhys2010,WeberPRB2010}.


We conclude by discussing the Fermi liquid cut-off energy scale $\omega_c$ which is expected to vanish with the quasiparticle residue $Z$~\cite{MiyakeSSP1989,ByczukNatPhys2007,JackoNatPhys2009}. 
For the simplest 
Fermi liquid with isotropic Im$\Sigma$, the residue is given by $Z=v_\mathrm{F}/v_b$. If Im$\Sigma$ in addition is monotonically decaying to zero above the cut-off energy $\omega_c$, then  $Z \propto   \omega_c/W$ where  $W$ is a bare 
energy scale~\cite{JackoNatPhys2009}.
Hence the cut-off energy $\omega_c$ is an indicator of electron-electron interaction strength. 
The bandwidth normalization enables comparison of different materials classes. However, heavy fermion systems in the limit $Z\rightarrow 0$ typically have $\omega_c$ far below the instrumental energy resolution. This concern is irrelevant for PLCCO as weak interactions manifest as a large Fermi-liquid cut-off energy scale.
As shown in Fig.~\ref{fig:fig2}(f), $\omega_c\sim \SI{0.4}{eV}$ ($\omega_c^2 \sim \SI{0.16}{eV^2}$) in PLCCO is twice as large as 
that of
the nodal region in overdoped LSCO~\cite{ChangNatComm13,FatuzzoPRB2014} (see Appendix~A for determination of $\omega_c$). 
However, in LSCO the self-energy is not isotropic
and for both LSCO and PLCCO, Im$\Sigma\propto \omega$ for $\omega >\omega_c$ (see Appendix~A). 
This implies that neither $Z=v_\mathrm{F}/v_b$ nor $Z \propto   \omega_c/W$ is 
expected to hold true.
Instead, a Kramers-Kronig transformation of Im$\Sigma$ suggests 
$Z \propto \mathrm{ln}^{-1} (W/ \omega_c)$ in the limit $\omega_c\rightarrow 0$~\cite{VarmaPRL1989}.
From a single ARPES spectrum, it is generally not possible to determine whether $Z$ is proportional to $\omega_c$ or $\mathrm{ln}^{-1} (W/ \omega_c)$. We therefore resort to fundamental Fermi liquid property underlying the Kadowaki-Woods relation. That is, quasiparticle lifetime and mass renormalization are expected to scale at least for comparable materials. Transport and thermodynamic experiments do support the Kadowaki-Woods relation~\cite{JackoNatPhys2009} though multi-band physics allows for numerous exceptional cases~\cite{hussey_non-generality_2005}. ARPES experiments have the advantage of extracting quasiparticle lifetime and mass renormalization not only from the same band but also from a very narrowly defined momentum region. As such, it makes sense to attempt construction of a spectroscopic version of the Kadowaki-Woods relation. In doing so, we here focus on transition-metal oxides with perovskite-based crystal structures ranging from pseudo-cubic (LaNiO$_3$)~\cite{EguchiPRB2009} to tetragonal (LSCO and PLCCO) and orthorhombic (\CSRO ) ones. We stress that the extremely strongly correlated regime, represented by U- and Ce- based heavy fermion systems, is expected to have $\omega_c \rightarrow 0$ falling below our energy resolution.

Within our selected material class, we are seeking a relation between the electron scattering factor $\beta$ and the quasiparticle mass renormalization factor $Z^{-1}$. 
In Fig.~\ref{fig:fig3}, we thus plot $W\beta$ versus $\mathrm{ln}^{2} (W/ \omega_c)$ with $W$ being a quarter of the DFT bandwidth~\cite{PavariniPRL2001,IkedaPRB2009,HamadaJPCS1993,HaverkortPRL2008} 
for PLCCO, LSCO, 
LaNiO$_3$~\cite{EguchiPRB2009} (\dx\ band~\cite{NowadnickPRB2015}) and \CSRO\ (\dxz/\dyz\  band).
The self-energy analysis of  \CSRO\ is presented in Appendix~A.
Combined, these correlated metals follow $W\beta  \propto \mathrm{ln}^{2} (W/ \omega_c)$ 
over more than 
an order of magnitude on both axes. 
This spectroscopic 
analogue of the Kadowaki-Woods relation suggests a logarithmic connection between the Fermi liquid cut-off energy scale and the quasiparticle mass renormalization factor $Z^{-1}$. The Fermi liquid properties (quasiparticle lifetime and mass) are thus set by a single energy scale; the Fermi liquid cut-off $\omega_c$ that smoothly connects weakly and strongly correlated Fermi liquids. 


\section{Conclusions}
In summary, we have carried out SX and VUV ARPES measurements on the electron-overdoped cuprate PLCCO. 
A two-dimensional electronic structure was revealed by SX ARPES experiments. This in turn enabled precise determination of PLCCO's in-plane self-energy using VUV light.
In contrast to the hole-doped counterpart LSCO, 
PLCCO displayed weak momentum-isotropic Fermi-liquid excitations. 
Characteristic parameters such as the scattering-rate coefficient $\beta$ and the Fermi-liquid cut-off energy $\omega_c$ revealed 
weak electron correlations 
compared to those reported in LSCO and \CSRO, but close to LaNiO$_3$.
Despite these strong contrasts, the 
four systems were found to satisfy a common relation that connects $\beta$ to $\omega_c^{-1}$, and hence to the mass renormalization factor $Z^{-1}$. 
Our results 
constitute a spectroscopic version of the Kadowaki-Woods relation $\beta  \propto Z^{-2}$. We reveal how this relation emerges from the quasiparticle lifetime and mass being set by a single 
energy scale $\omega_c$ that characterizes all Fermi liquids.

\section{Appendix A: Supplemental self-energy data of PLCCO, LSCO and \boldmath$\mathrm{Ca}_{1.8} \mathrm{Sr}_{0.2} \mathrm{RuO}_4$}

	
 
\begin{figure}[ht]
 	\begin{center}
 		\includegraphics[width=0.48\textwidth]{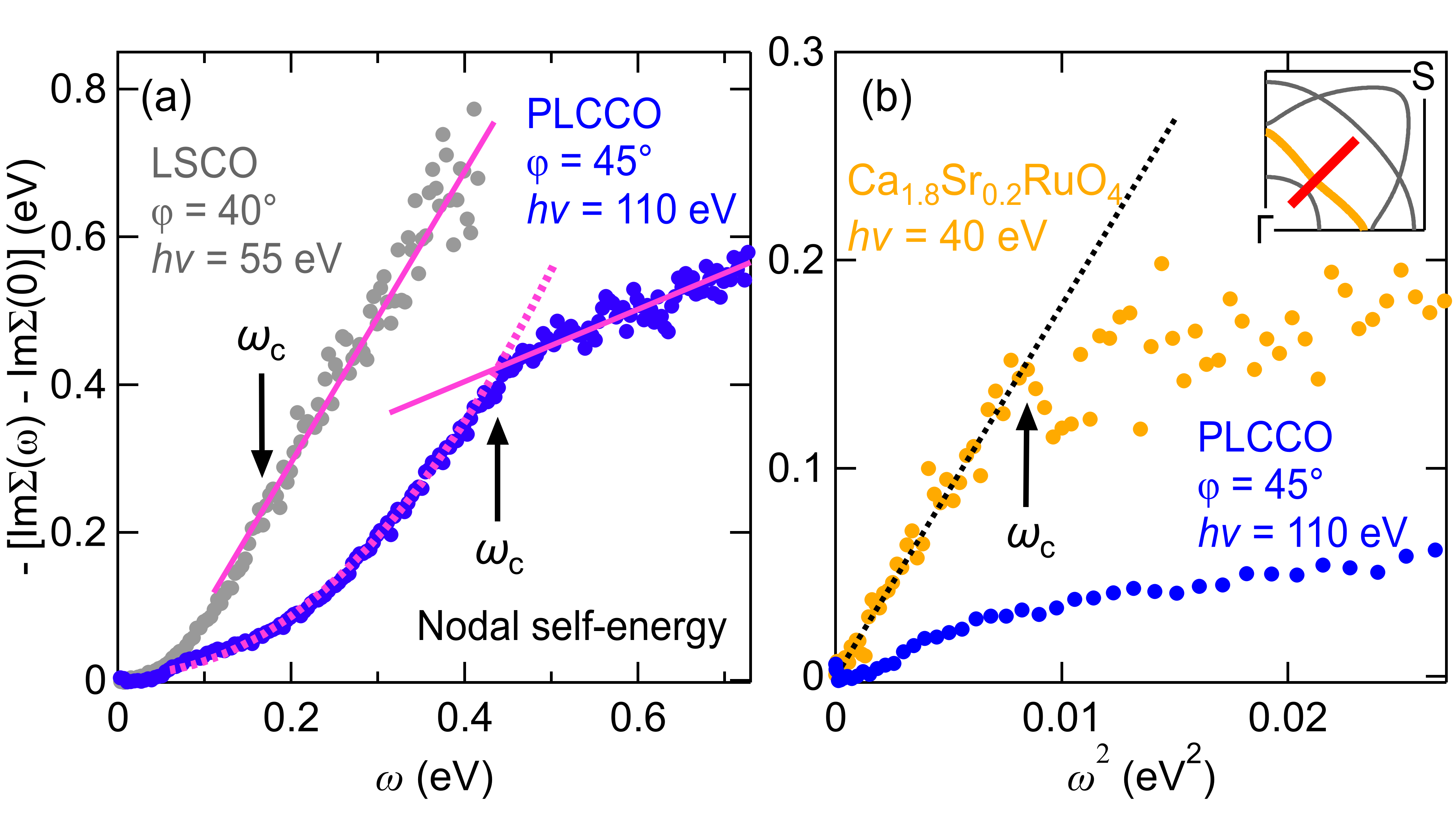}
 	\end{center}
 	\caption{\textbf{Self-energy -Im$\Sigma$($\omega$) of PLCCO, LSCO and \boldmath$\mathrm{Ca}_{1.8} \mathrm{Sr}_{0.2} \mathrm{RuO}_4$.} (a) Nodal self-energy -Im$\Sigma$($\omega$) of PLCCO and LSCO plotted as a function of $\omega$. The LSCO data are identical to those plotted in Fig.~\ref{fig:fig2}(f) \cite{ChangNatComm13}. The PLCCO data were recorded at $h\nu = 110$~eV. An $\omega$-linear dependence is found for both compounds above the cut-off energy $\omega_c$, as indicated by solid lines. The $\omega^2$ dependence below $\omega_c$ is indicated by a dotted curve. (b) Self-energy -Im$\Sigma$($\omega$) of \CSRO\ plotted as a function of $\omega^2$. The orange Fermi surface sheet in the schematic inset is measured along the red cut and analyzed. The Fermi-liquid cut-off is marked by a black arrow. The nodal self-energy of PLCCO in panel (a) is re-plotted for comparison.} 
	
	 	\label{fig:figSuppl}
 \end{figure}

In Fig.~\ref{fig:figSuppl}(a), we plot as a function of $\omega$ the nodal self-energy -Im$\Sigma$($\omega$) of PLCCO and LSCO. The LSCO data are identical to that in Fig.~\ref{fig:fig2}(f). For PLCCO, data recorded down to deeper binding energies using $h\nu = 110$~eV photons are plotted. In contrast to the low-energy part that scales with $\omega^2$ [see also Fig.~\ref{fig:fig2}(f)], the self-energy above the cut-off $\omega_c$ exhibits an $\omega$-linear dependence for both compounds. For the PLCCO data in Figs.~\ref{fig:fig2}(f) and \ref{fig:figSuppl}(a), low-energy (0.06~eV $< \omega <$ 0.4~eV) parabolic and high-energy ($\omega >$ 0.45~eV) linear $\omega$ dependences are respectively extrapolated to find a crossing point, which is defined as $\omega_c$.


 

ARPES experiments on \CSRO\ were carried out at the 1$^3$ beamline of BESSY. The sample was cleaved \textit{in-situ} under ultra-high vacuum and measured at $T=30$~K with $h\nu=40$~eV incident light. 
The orange Fermi surface sheet in the schematic inset of Fig.~\ref{fig:figSuppl}(b), which is of $d_{xz}/d_{yz}$ character~\cite{SutterPRB2019}, is measured along the red cut and analyzed. From momentum-distribution-curve peak width, the inelastic part of -Im$\Sigma$($\omega$) is constructed and plotted in Fig.~\ref{fig:figSuppl}(b) as a function of $\omega^2$. Here, the bare-band Fermi velocity of $v_F = 2.34$~eV~\AA\ derived from a DFT calculation~\cite{SutterPRB2019} is used. The Fermi-liquid cut-off, indicated by a black arrow, is defined as the crossing of extrapolated low-energy ($\omega < 0.08$~eV) parabolic and high-energy ($\omega > 0.11$~eV) linear $\omega$ dependence. 

\section{Acknowledgments}
M.~Horio, K.P.K., Q.W., K.v.A., and J.C. acknowledge support by the Swiss National Science Foundation. C.E.M acknowledges support from
the Swiss National Science Foundation under grant No.~P400P2\_183890. Y.S. is funded by the Swedish Research Council (VR) with a Starting Grant (Dnr.~2017-05078) as well as Chalmers Area Of Advance-Materials Science. A.F. is funded by JSPS KAKENHI Grant No.~JP19K03741. We thank DESY (Hamburg, Germany), a member of the Helmholtz Association HGF, for the provision of experimental facilities. Parts of this research were carried out at PETRA III. Funding for the ASPHERE III instrument at beamline P04 (Contracts 05KS7FK2, 05K10FK1, 05K12FK1, and 05K13FK1 with Kiel University; 05KS7WW1 and 05K10WW2 with W\"urzburg University) by the Federal Ministry of Education and Research (BMBF) is gratefully acknowledged. ARPES measurements were carried out at the SIS and P04 beamlines of the Swiss Light Source and DESY, respectively.



%

\end{document}